# Degradation of crude oil and pure hydrocarbon fractions by some wild bacterial and fungal species

Srwa A. Mohammed*[1]; Taha J. Omar[1]; Ayad H. Hasan[1,2]

[1] *Dept. of Medical Microbiology, Faculty of Science and Health, Koya University, Koya KOY45, Kurdistan Region - F.R. Iraq*

[2] *Department of Biomedical Sciences, College of Health Technology, Cihan University-Erbil, Kurdistan Region, Iraq.*

* *Corresponding author: Srwa A. Mohammed (srwa.ali@koyauniversity.org)*

**Introduction:** petroleum hydrocarbons are a major concern due to their widespread distribution into the environment, such as soil or water, and their harmful effects on humans. (Chen *et al*., 2015; Wang *et al.,* 2018). Biological processes are utilised in the bioremediation process, which is a collection of technologies that either aid in the elimination of contaminants or make them minimally hazardous (Silva *et al.,* 2015). **Objectives:** This study was conducted to isolate and identify bacterial and fungal species from the soil of the TaqTaq (TTOPCO) oil field in the Kurdistan Region of Iraq, and then investigate their ability to degrade crude oil and its fractions. Many studies agree with the present study for the biodegradation of crude oil *in vitro* by using different methods and various concentrations of crude oil (Ramdass and Rampersad, 2021).

## Abstract

The use of biodegradation as a method for cleaning up soil that has been contaminated by spilt petroleum can be an effective strategy. So, this study investigated the existence of the wild microorganism in soil contaminated with oil and study their ability to degrade petroleum *in vitro*. Nineteen samples were collected from various locations near Taq Taq (TTOPCO) natural seeps in the Kurdistan Region of Iraq. Morphological, cultural, biochemical tests and molecular identification were used to identify the microbial communities, in addition, spore texture and the colour of the fungal isolates were investigated on the fungal isolates. Out of the19 samples, 17 indigenous bacterial strains and 5 fungal strains were successfully isolated. From the absorption spectrophotometry, *Bacillus anthracis, Bacillus cereus*, *Achromobacter* sp. and *Pseudomonas aeruginosa* for the bacterial isolates grew well on a minimal salt medium supplemented with 1% crude oil. Results showed that these isolates mentioned above had a strong ability to degrade crude oil by reducing the colour of 2,6-dichlorophenol indophenol (DCPIP) from deep blue to colourless. However, for the fractions of hydrocarbon, the bacterial isolates failed and did not affect the colour of any of the fractions. The results for fungi showed that *Aspergillus lentulus* and *Rhizopus arrhizus* had a strong ability to degrade both crude oil and fraction F1 by reducing the colour of DCPIP. Each fungal isolates also had a great tolerance to different concentrations of crude oil when grown on solid MSM. This study showed these microorganisms have a strong ability to degrade crude oil and can be used to clean up soil and the environment.

## 1. Introduction

The Kurdistan Region of Iraq (KRI) is located north and northeast of the Arabian plate. The region is one of the oil-rich areas of Iraq Shlimon, *et al.,* (2020) Koya city is one of the oil-rich areas in the Kurdistan Region with many reservoirs. TaqTaq (TTOPCO) reservoir is one of them. The intensive use of petroleum, however, results in environmental disruption Xue *et al*. (2015). Spills that occur during and/or as a result of petroleum extraction, storage, refining, manufacturing, shipping, oilfield development, leakage from oil pipelines or tankers, and discharges of petroleum hydrocarbons are also major concerns due to their widespread distribution into the environment, such as soil or water, which affects humans' health (Chen *et al*., 2015; Wang *et al.,* 2018). Crude oil is a complex blend of hydrophobic components such as n-alkanes, aromatics, resins, and asphaltenes Barnes, *et al*. (2018).

Managing hydrocarbon contamination has become easier than before due to the development of several new technologies in recent years. Biological processes are utilised in the bioremediation process, which is a collection of technologies that either aid in the elimination of contaminants or make them minimally hazardous (Silva *et al.,* 2015). The procedure is cost-effective and can be applied in its entirety to the area that is contaminated. Consequently, microbial remediation is a promising method for the complete mineralisation of hydrocarbons into carbon dioxide and water (Wang *et al*., 2015).

Bacteria, fungi, and yeast biodegrade hydrocarbons in the environment. Some bacterial species can metabolise specific alkanes others break down aromatic or resin fractions of hydrocarbons in many different manners depending on the oxygenase (Xu, *et al.* 2017).

Although several fungi can grow in soil, few species can survive in contaminated soils with biodegradation efficiency ranging from 6% to 82% (Juhasz and Naidu, 2000; Das and Chandran, 2011; Acevedo *et al*., 2012). This study aimed to isolate and identify bacterial and fungal species from the soil of the TaqTaq (TTOPCO) oil field and investigate their ability to degrade crude oil and its fractions using spectrophotometry and 2,6-dichlorophenol indophenol (DCPIP) methods and mycelial radial growth measurements.

## 2. Material and Method

### 2.1. Sample and sampling locations

All the samples were collected from the Koya City Taq Taq asphalt seep (TTOPCO), which is in the Kurdistan Region of Iraq (KRI) Five samples of contaminated soil were collected at a depth of 5 to 10 cm from each of the following four different locations: 1- Mud site (group A) the samples were given the numbers 1–4, it was oil and water mixed samples were collected from an area close to the drilled pool of oil at the Taq Taq oil seep. 2- Underlying and flanking region (group B) of the Taq Taq asphalt seep flow the samples were given the numbers 5–9. 3- Ten meters away from site number two, it had not been contaminated by any oil spills, and it was used as a control for analysing the contaminated soil, the samples in this group were given numbers 10–14. 4-The transportation area where the soil was contaminated with spilt oil, was 20 m away from site number two, the collected samples were numbered 15–19. All the samples were placed into sterilised polyethylene bags. Approximately 2 L of oil samples were collected from operating oil wells and stored in bottles that had been carefully sealed. The bottles and polyethylene bags were placed in a container packed with ice until then transported to a laboratory and stored at 4°C.

2.2. Morphological and molecular identification of isolated bacteria and fungi strains

Pure culture technique was applied to isolate pure single colonies from each soil sample using nutrient agar, McConkey agar, cetrimide agar and mannitol salt agar (Ozyurek and Bilkay, 2017). Then, biochemical tests and gram stain were conducted for all the isolated single colonies (Riedel, *et al*. 2019). The fungi were identified using morphological and taxonomic keys found in mycological keys (Watanabe 2018).

The genomic DNA of the isolated bacteria and fungi was extracted by QIAwave DNA Blood and Tissue Kit (Germany, Cat. no. 69556) according to the manufacturer's protocol. The Nanodrop (NanoDrop Spectro 117 432-UK) spectrophotometer was used to determine the quality and quantity of the genomic DNA. 16S rRNA gene for bacteria that contain a highly variable region was amplified by the PCR (Polymerase Chain Reaction) using the universal primer EubA F (AAGGAGGTGATCCANCCRCA) and EubB R (AGAGTTTGATCMTGGCTCAG), which will give an end product size of 1534 bp Lane, (1991). On the other hand, the internal transcribed spacer (ITS) region for fungi was amplified using LROR (ACCCGCTGAACTTAAGC.) and LR6 CGCCAGTTCTGCTTACC Raja, *et al*. (2017), which will give an end product size of 1200 bp (Macrogen inc, South Korea company).

Bacterial amplification reactions were performed in a final volume of 25 µl of PCR reaction mixture using Prime Q5 Hot Start High-Fidelity 2X Master Mix (Cat. no.M0494S). The PCR reaction included 12.5µl of 2x Master Mix, 5 pmol (1 µl) of each of forward (Eub A) and reverse (Eub B) primers, 100 ng (1 µl) template DNA, and 9.5 µl nuclease-free water. The PCR process was performed using BIO-RAD T100TM Thermal Cycler (UK) and programmed as follows: 2 min of initial denaturation at 98°C, followed by 25 cycles of reaction with the 50s of denaturing at 98°C, 50s of annealing at (64.3, 61, 59, 58, and 56) °C, 50s of extension at 72°C, and the final extension was performed for 4 min at 72°C.

To amplify the ITS region of the fungal isolates the PCR reactions were performed in a final volume of 25 µl of the reaction mixture, which included 12.5µl of 2x Master Mix, 5 pmol (1.5 µl) of each of forward (LROR) and reverse (LR6) primer, XX ng (3 µl) template DNA, and 6.5 µl nuclease-free water. The PCR program procedure was carried out as previously described for bacterial isolates with an annealing temperature of 56°C and a cycling repetition of 34 cycles.

The efficiency of DNA extraction was evaluated using electrophoresis on a 1% agarose gel (w/v) that was stained with ethidium bromide (0.5 g/ml) and had a 100bp DNA marker (DENA 100 bp plus DNA size marker II S-5091). The gel was run at 80 V for 1:30 hours in a 1X TBE buffer. After the experiment had run its course, ultraviolet light was used to visualize the DNA bands, and photographs were taken using UV Gel Imager SynGene 1409 Lee, *et al.* (2012).

All the PCR amplicons were sent out for sequencing by Macrogen inc, a South Korean company. Sequence quality, analysis, and editing were carried out using the DNA baser assembler tool. The 16S rDNA sequence was compared to previously identified bacterial DNA sequences using the BLASTN in order to classify the bacterial isolates independently (http://www.ncbi.nih.gov/BLAST).

2.3. Crude oil and pure hydrocarbons degradation by bacteria and fungi isolates

A minimal Salt Medium (MSM) was used for bacterial isolates (Sigma-Aldrich). 1ml/L trace element ($CaCl_{2.2}H_2O$ 4.77g/100ml, $FeSo_4.7H_2O$ 0.37 gm/100ml, $MnCl_2.4H_2O$ 0.10 gm/100ml, $Na_2MoO_4.2H_2O$ 0.02 gm/100ml at pH 7) was added to the MSM, in addition to 1 ml/L vitamin mix solution ( Pyridoxine-Hcl 10.0 mg, p-Aminobenzoic acid 5.0 mg, Lipoic acid 5.0 mg, Nicotinic acid 5.0mg, Riboflavin acid 5.0mg, Thiamine-Hcl 5.0mg, Calcium D1-Pantothenate 5.0mg, Biotin 2.0mg, Folic acid 2.0mg, Vitamin B12 0.1mg) and 0.1gm/L yeast extract.

Minimal salt medium (MSM) and potato dextrose broth were used for the isolation and maintenance of fungal isolates. Five grams of each collected soil sample were incubated in 250 Erlenmeyer flasks containing 100ml freshly prepared MSM containing NaCl (0.5g/l), (NH4)2SO4 (0.1g/L), NaNO3 (0.2 g/L), MgSO4.7H2O (0.025 g/L), K2HPO4.3H2O (1g/L) and KH2PO4 (0.4g/L) at pH 7.0.

Crude oil was fractionated using simple distillation by pouring it into the round bottom flask and adjusting the fractionation column and thermometer with the conical flask on one side and the condenser on the other side. The temperature was measured with a thermometer, and four fractions of crude oil were separated. The temperature degree of fraction number was (40-60)°C belong to F1, (60-80) °C belong to F2, (80-100)°C belong to F3 and (100-130) °C belong to F4, respectively. Consequently, degradation of all the fractions was conducted as mentioned above using 1% of each fraction for bacterial and fungi isolates.

Experiments on biodegradation were carried out in glass screw cup tubes containing 10 ml of MSM and 1% crude oil as the sole carbon source. Prior to adding crude oil, both the media and the crude oil were sterilized separately by autoclaving at 121°C for 15 minutes. A single colony of the isolate was inoculated into 10 ml nutrient broth for bacteria and incubated overnight at 30°C at 150 rpm. Following incubation, the culture was centrifuged for 10 minutes at 10000 rpm. The bacterial suspension was ready for use after the cell pellets were washed to remove all nutrients and re-suspended in normal saline until the OD at 600 nm was equivalent to 1. 1% of the bacterial suspension was transferred into the MSM supplemented with crude oil. Uninoculated media was used as a control, and all tests and controls were done in triplicate. The same procedure was followed for pure hydrocarbon fractions (F1, F2, and F4). Two methods were used to detect crude oil and hydrocarbon fractions degradation: 1- spectrophotometer method: the absorbency at wavelength 600 nm was measured at 24 hours, 48 hours, one week, three weeks, and eight weeks to determine the bacterial growth. 2- 2,6-Dichlorophenol indophenol (DCPIP) method was used to evaluate bacterial isolates' ability to degrade crude oil and hydrocarbon fractions. 1 ml washed bacterial cells (Optical density (OD) at 600 nm was equivalent to 1), 1% V/V crude oil and each fraction separately, and 1% (0.6 mg/L) of redox indicator (DCPIP) were added to 10 ml MSM, and incubated for two weeks in a shaker incubator at 30°C with 150rpm. Then, every 24 hrs, the colour changes are monitored. Uninoculated media was used as a control (Selvakumar, 2014; Balogun, 2015). Likewise, this method was repeated for fungal isolates.

The fungal mycelia were grown in a glass screw cup container with 10 mL of mineral salts medium supplemented with 1% crude oil and 1% 2,6-Dichlorophenol indophenol (DCPIP) redox as followed in bacterial degradation for 7 days at 30°C and 150 rpm in shaker incubator without any other nutrient as additive. crude oil as a carbon source was used Barnes, (2018).

Measuring mycelial radial growth is the second method used to detect crude oil and pure hydrocarbons degradation with fungi by the cultivation of all fungal isolates on solid MSM media supplemented with different concentrations (5, 10, 15, 20) % of crude oil and 0% was used as a control. All of the plates were incubated for 7 days at 30°C. Thus, the mycelial radial growth cm/day was measured (Anaisell, *et al.,* 2014).

## 3. Result and Discussion

3.1. Phenotypic and molecular characterization of bacterial and fungal isolates

17 out of 19 bacterial samples could be recognized by utilizing Bergey's manual of systematic bacteriology for phenotypic and colony identification Ozyurek (2017), while the other two samples showed no signs of growth. In order to confirm the identity of the samples, total DNA was extracted from all the seventeen bacterial isolates and then amplified using EubA F and EubB R primer pair that bind specifically to the 16S rRNA gene as mentioned in section 2.2. The expected size of the DNA fragment (1534 bp) was amplified successfully. No PCR products were detected in the negative control as shown in (Figure 1).

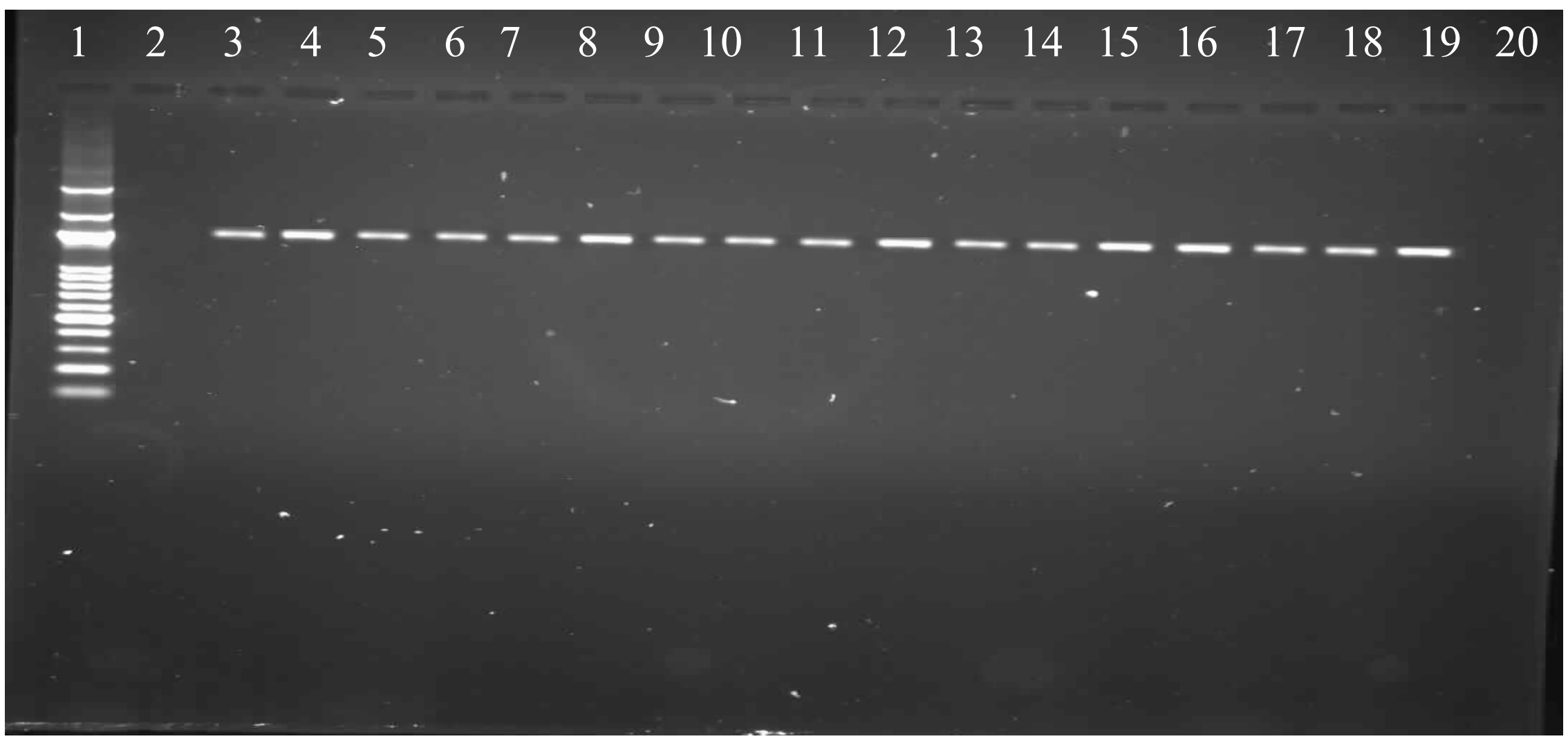

**Fig. 1.** Partial amplification of 16S rDNA using EubA and EubB primers. Lanes 1 and 2 represent a 100bp DNA marker (DENA 100 bp plus DNA size marker II S-5091), a negative control that has been run without a DNA template, respectively. Lanes 3 through 19 shows ~1534bp of PCR amplicons generated using a DNA template from number 1 through number 17 bacterial samples, respectively.

The result showed that the following microorganisms were isolated from group A soil samples after the classical identification and molecular confirmation. C*aldibacillus thermoamylovorans* strain SSBM chromosome with accession number CP023704, *Bacillus pumilus* strain mv49b KU230016, *Bac*illus *tropicus* strain NP_ 2. OP048825.1, and *Pseudomonas aeruginosa* strain DM Bust3A MF599526.1, respectively. Group B soil samples included *Aneurinibacillus migulanus* strain DSM 2895 with accession number NR 112214.1, *Achromobacter sp.* MT093185.1, *Bacillus anthracis* strain FDAARGOS 695 with accession number CP054816.1, *Bacillus cereus* strain T11-12 with accession number HQ333011.1, and *Lysinibacillus sp*. 381 with accession number KT034471.1, respectively. Group C soil samples revealed *Paenibacillus dentritiformis* strain PV3-16 with accession number MH472941.1, *Aneurinibacillus migulanus* strain RD with accession number KX083693.1, *Brevibacillus borstelensis* strain ML13. with accession number MN604049.1, and *Bacillus paramycoides* strain EFBC 17 with accession number MN793201.1, respectively. Group D soil samples contained *Bacillus anthracis* strain FDAARGOS_695 with accession number CP054816.1, *Pseudomonas stutzeri* strain HA549 with accession number KJ535356., *Bacillus paramycoid* 2883 with accession number MT611845.1*,* and *Lysinibacillus capsici* strain anQ-h6 with accession number CP084108.1, respectively. Except for *Bacillus anthracis* and *Bacillus cereus*, most *Bacillus spp.* are not pathogenic, and many species have been exploited for biotechnological and industrial uses Gu, *et al*. (2019).

Above results are in agreement with the results reported by Chen *et al.* (2017) who isolated *Exiguobacterium sp. Pseudomonas aeruginosa*, *Alcaligenes sp*., and Bacillus sp., petroleum hydrocarbon-degrading microorganisms' area contaminated with petroleum**.** And also, the study of **(**Szczepaniak *et al*., 2015) demonstrated Petroleum hydrocarbons are compounds which undergo decomposition in soil due to the activity of several groups of microorganisms. Several different microbial species participate in the biodegradation of hydrocarbons, ranging from strictly aerobic to strict anaerobic bacteria. Several Gram-positive (Rhodococcus or Bacillus) as well as Gram-negative (Alcaligenes, Acinetobacter, Pseudomonas) species are also characterized by a relatively broad substrate spectrum.

Five fungi isolates were isolated from the 19 soil samples. The expected size of PCR amplicons (1200 bp) was detected successfully in the all five samples. No PCR products were detected in the negative control as shown in (Figure 2). Therefore, to track down each fungal isolate to its exact species, the PCR amplicons of each isolate were sent out for sequencing using the forward (LROR). Thus, the isolate which was isolated from group A was *Aspergillus lentulus* 28S ribosomal RNA with accession number XR 004500616.1, and the isolates which were isolated from group B were *Aspergillus fellis* strain FM324 chromosome 3. with accession number CP066505.1, *Aspergillus luteonubrus* strain MST FP2246 with accession number MT196912.1 and *Aspergillus arizonicus* isolate CCF 5341 with accession number OK321187.1, respectively. The isolate which was isolated from group C was *Rhizopus arrhizus* Strain SC49B03 with accession number MW113537.1. The study of (Barnes *et al.,* 2018) demonstrated that the ten fungal isolates chosen were able to grow and degrade crude oil. The isolate *Penicillium citrinum* NIOSN-M126 isolated from Divar mangrove sediments demonstrated the greatest ability to utilize crude oil, followed by *Aspergillus flavus* NIOSN-SK56S22 isolated from Arabian Sea sediments.

3.3. Degradation of crude oil and crude oil fractions by bacterial isolates and fungi isolates:

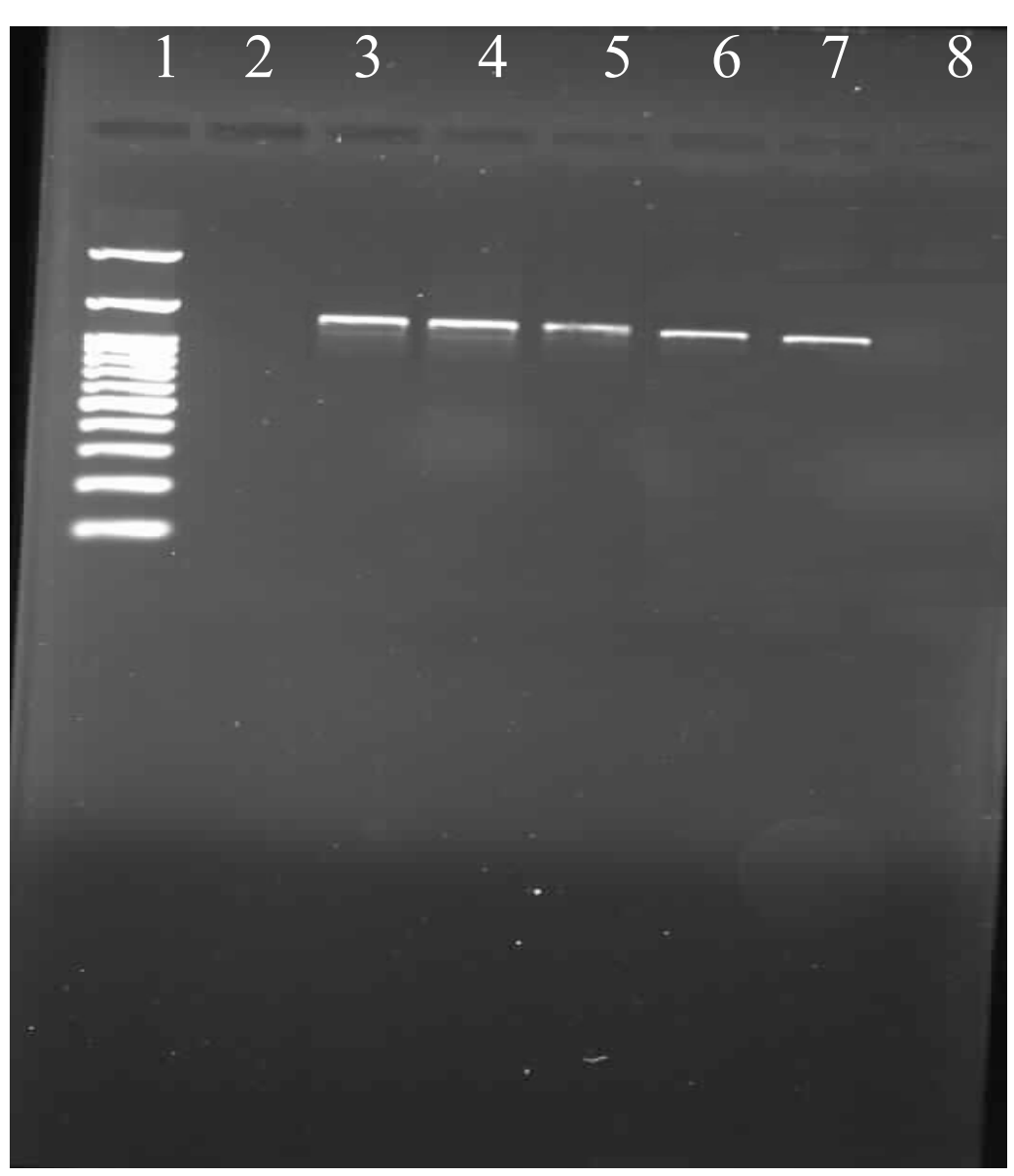

**Fig. 2.** Partial amplification of ITS using (LROR)/(LR6) primers. Lanes 1 and 2 represent a 100bp DNA marker (DENA 100 bp plus DNA size marker II S-5091), a negative control that has been run without a DNA template, respectively. Lanes 3 through 7 shows ~1200 bp of PCR amplicons generated using a DNA template from number 1 through 5 fungal samples, respectively.

Bacteria: strains were isolated for their capacity to grow in the presence of crude oil and the fractions of crude oil used in the investigation for detecting which bacterial isolates have a high ability to degrade crude oil and the fractions of crude oil. Two methods were used for this purpose.

spectrophotometer method: The results showed that growth was determined through measuring the absorbency at wavelength 600 nm by adding 1% washed bacterial cell reading OD at 600 nm was equivalent to 1 and 1% crude oil and incubated at 30°C in a shaker incubator 150rpm. Then, the results showed that no. (3, 4, 5, 6, 7, 8, 9, 10,11, 12, 13,14, and 16) the OD reading after 24 hrs were (0.50, 0.43, 0.45, 0.68, 0.61, o.49, 0.60, 0.52, 0.68, 0.70, 0.46, 0.73, and 0.60) respectively. However, the OD reading after 48 hrs was increased in some of the sample's no. (4, 6, 7, 8, 11, 12, and 14) the OD reading at 600 nm was (0.56, 0.32, 0.69, 0.57, 0.32, 0.79, and 0.92) respectively. Meanwhile, the OD reading was decreased in other samples. Then, the results were shown after one week the samples number (4, 6, 7, 8, 11, 12, and 14) was recorded OD reading (0.283, 0.299, 0.4271, 0.3544, .2947, 0.3253, and 0.5793) respectively. Furthermore, after three weeks number (6, 7, 8, 12 and 14) OD reading was (0.21, 0.34, 0.24, 0.21, and 0.43) respectively. Finally, after 8 weeks all the isolates decreased due to decreasing carbon sources and nutrients as shown in Table 1. Therefore, this investigation concluded that the most potent isolates for resisting crude oil and growing well in minimal media supplemented by crude oil were (*Pseudomonas aeruginosa*, *Achromobacter sp. and Bacillus anthracis*, *Bacillus cereus, Aneurinibacillus migulanus,* and *Brevibacillus borstelensis*) and (Figure 1) showed bacterial colony on solid MSM.

Degradation hydrocarbon fractions (F1, F2, F3, and F4) were managed by a simple distiller at the laboratory and to detect hydrocarbon degradation spectrophotometer method was used for the samples which degrade crude oil faster than others. Then, the number (4, 6,7, 8,11,12,14, 16 and 17) were undergone to this test. And the results showed that (F2 and F4) were consumed by the isolates more than other fractions. The OD reading for isolates that consume and grow well in F2 were (4,6, 7,12,14, and 17) the OD reading were (0.312, 0.334, 0.456, 0.594, and 0.482) respectively, Furthermore, numbers (8,11 and 17) were consumed fraction 4 (F4) more than the other fractions, the OD reading were (0.357, 0.3320. and 0.334) respectively. The bacterial growth was increased until after four weeks decrease the OD reading and growth of bacteria for all the isolates were caused by decreasing the carbon source in the media. Yu *et al*. (2005b) showed in their study that Autochthonous microorganisms in sediments also possessed satisfactory Polycyclic aromatic hydrocarbons (PAHs) degradation capability and all three PAH were completely degraded after 4 weeks of growth.

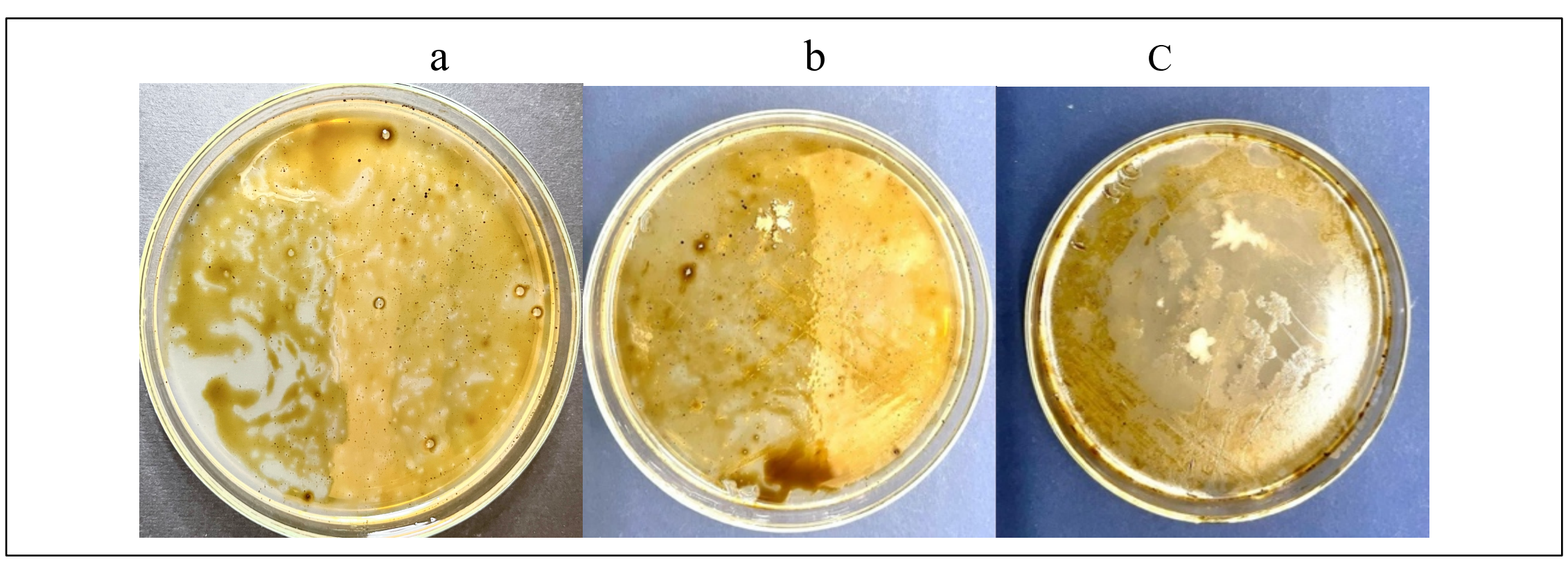

**Fig. 3.** Bacterial isolates that grown on MSM supplemented with 1% crude oil a- *Bacillus cereus* b- *Achromobacter sp.* c- *Pseudomonas sp*

A 2,6-Dichlorophenol indophenol (DCPIP) method: This method also was used to assess the ability of bacterial isolates to degrade crude oil. The suspension was prepared as mentioned in section 2.2. When compared to the control, the change in colour of the inoculation degradation medium from deep blue to colourless demonstrated the ability of the bacterial isolates to degrade crude oil. Then, the results showed that numbers (7, 8, 11, 12, and 14) have a high ability to reduce the colour completely to colourless after 4 days, and numbers (4, 6, and 16) also can reduce the colour but less than the numbers mentioned above as shown in (Figure 2). Many studies agree with the present study for the biodegradation of crude oil in vitro by using different methods and different concentrations of crude oil. Ramdass and Rampersad, (2021) showed crude oil degradation by bacteria, fungi and yeast and they were concluded 2% of crude oil were mixed with two media types. the result indicated that all microbes recovered were able to utilize crude oil on both media types in vitro. Rizi *et al*. (2012) reported gram-positive bacteria such as *Bacillus cereus* and *Bacillus subtilis* were predominant in degrading crude oil this paralleled with the present study that *Bacillus cereus* and *Bacillus spp.* the most predominant in degrading crude oil. while, the results showed the colour does not change for all the fractions (F1, F2, F3, and F4) separately for bacterial isolates.

**Table 1.** Spectrophotometer readings for bacterial isolates that cause high turbidity in MSM when 1% crude oil is added

| **No. of isolates** | **After 24 hrs** | **After 48 hrs** | **After one week** | **After 3 weeks** | **After 8 weeks** |
|---|---|---|---|---|---|
| 3 | 0.5028 | 0.231 | 0.0594 | 0.0887 | 0.0001 |
| 4 | 0.4352 | 0.564 | 0.2830 | 0.0568 | 0.0002 |
| 5 | 0.4591 | 0.152 | 0.2514 | 0.0343 | 0.0001 |
| 6 | 0.6872 | 0.327 | 0.299 | 0.2110 | 0.0003 |
| 7 | 0.6123 | 0.695 | 0.4271 | 0.3454 | 0.0002 |
| 8 | 0.4973 | 0.577 | 0.3544 | 0.2499 | 0.0001 |
| 9 | 0.6069 | 0.210 | 0.0931 | 0.0903 | 0.0002 |
| 10 | 0.5265 | 0.257 | 0.1639 | 0.1259 | 0.0001 |
| 11 | 0.687 | 0.325 | 0.2947 | 0.2054 | 0.0001 |
| 12 | 0.7090 | 0.796 | 0.3253 | 0.2106 | 0.0002 |
| 13 | 0.4686 | 0.481 | 0.0553 | 0.0319 | 0.0002 |
| 14 | 0.7327 | 0.921 | 0.5793 | 0.4383 | 0.0002 |
| 16 | 0.6052 | 0.495 | 0.1906 | 0.1179 | 0.0001 |
| 17 | 0.3812 | 0.261 | 0.1232 | 0.0944 | 0.0001 |

Fungi: Five oil-degrading fungi were isolated from 19 samples of crude oil-polluted soil and control samples. Two methods were used to detect crude oil degradation by fungi isolates.

**17 16 15 14 13. 12 11 10 9 8 7 6. 5 4 3 2 1 C**

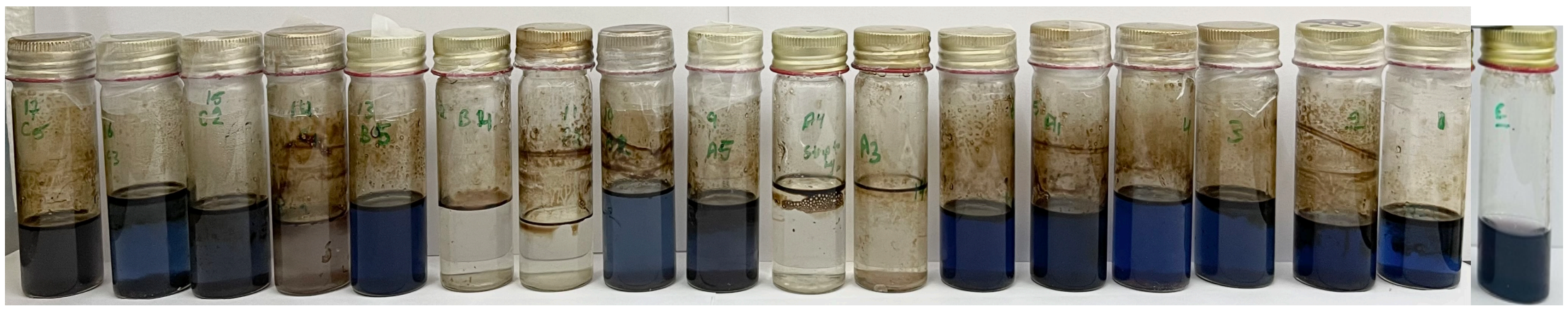

**Fig. 4.** Ability of the bacterial isolates which start from control © followed by number 1 through 17 to reduce 2,6-Dichlorophenol indophenol (DCPIP) by changing the colour from blue to colourless after 4 days

Mycelia radial growth measurement on minimal salt medium (MSM): Five isolated fungi were screened for their ability to tolerate crude oil or petroleum hydrocarbons from 0 to 20% concentration. All the isolated fungi showed to have a high ability to tolerate different crude oil concentrations as well as in the control treatment (without crude oil). *Rhizopus arrhizus* exhibited the highest growth rate, 6.8 cm/day at 0%, as well as (6.6, 6.6, 6.4 6.1) cm/day for crude oil concentrations of (5,10,15, and 20%). *Aspergillus lentulus* presented the second highest growth rate (5.3, 5.0, 4.4, 4.1 and 4.0) cm/ day, followed by other isolates. The experiment proved that all five strains were capable to grow in crude oil by using crude oil as the sole carbon source. (Al-Zaban, *et al.,* 2021) who isolated four fungi isolates and measured mycelia radial growth on minimal medium (MM) supplemented with crude oil at different concentrations. All isolates grew faster in the control treatment (without crude oil) than in the other treatments supplemented with varying concentrations of crude oil. However, he observed remarkable adaptation and were able to survive high concentrations of crude oil of up to 20%. Reyes-César, *et al.* (2014) results showed the high biodegradation potential of the *Talaromyces spectabilis* CCS12 strain and it can be used in the development of novel green processes due to its strong ability to metabolize PAHs in soil. While the present study showed *Rhizopus arrhizus* which isolate from a control sample has high biodegradation potential on MSM solid media added with up to 20% of crude oil and 1% of four fractions separately after three days of incubation and this novel and a new record for this genus and followed by isolating *Aspergillus lentulus.*

The 2,6-Dichlorophenol Indophenol (DCPIP) method was used to evaluate the ability of selected fungal strains to degrade crude oil and the hydrocarbon fractions; the results showed that *Rhizopus arrhizus. and Aspergillus lentulus.* have a strong ability to degrade crude oil and reduce the colour from deep blue to colourless and consequently, degrade the F1 fraction as well. While other fungi isolates can degrade crude oil and change the blue colour to colourless. However, the (DCPIP) colour does not change and remains the blue colour for the fractions when compared with the control. Al-Hawash, *et al*. (2018) demonstrated that two strains of *Penicillium sp.* RMA1 and RMA2 were isolated from the Rumaila oil field and can change the colour gradually from deep blue to colourless, and this reaction suggested that *Penicillium sp.* RMA1 and RMA2 could degrade crude oil. Environmental pollution caused by oil spillage is one of the century's major issues that must be resolved. Biodegradation can emerge or limit this problem.

## 4. Conclusion

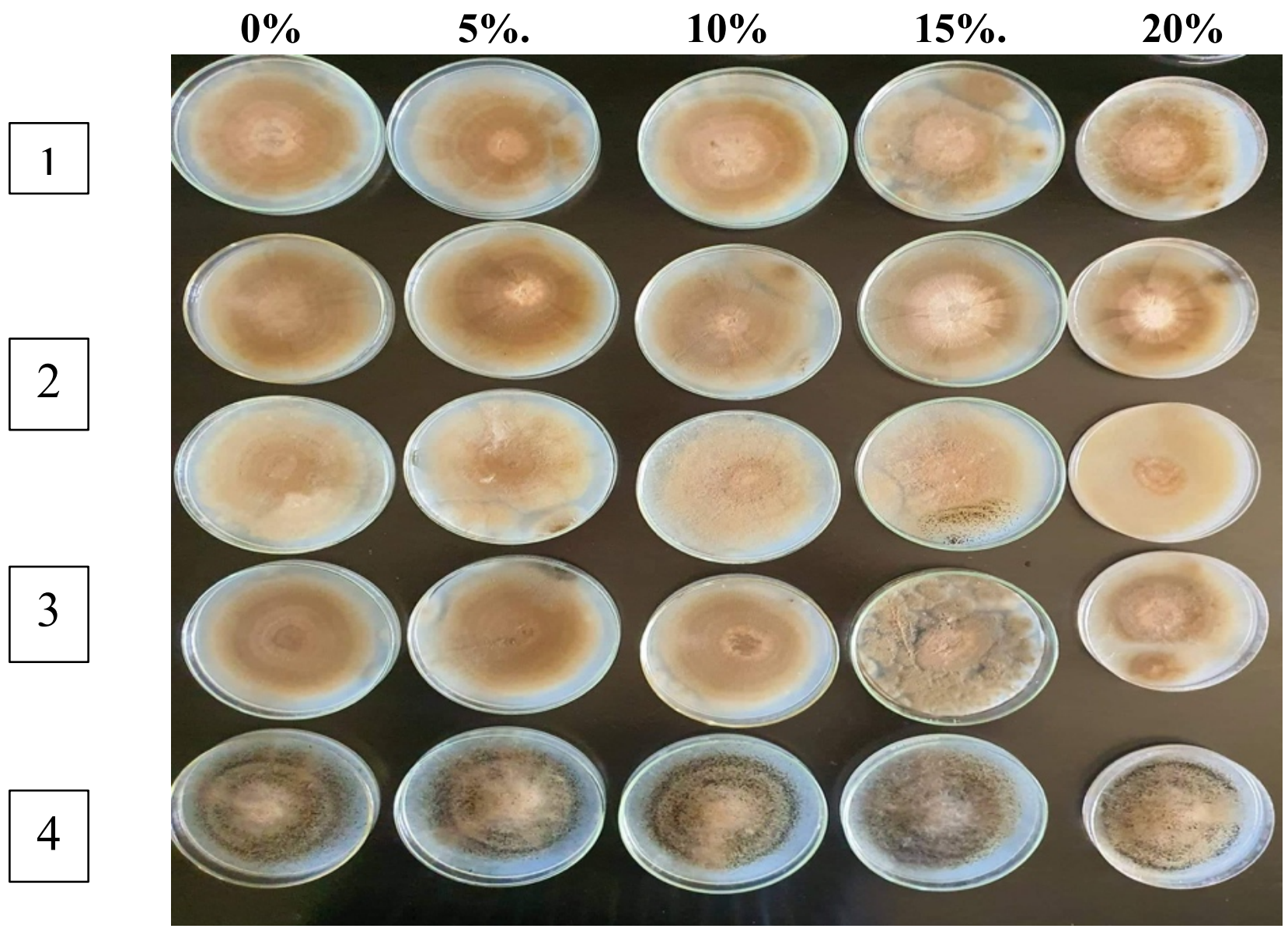

**Fig. 5.** Macroscopic images of radial growth of isolated fungal strains, in Petri dishes with different concentrations of crude oil (0% to 20%). Line 1 is *Aspergillus lentulus*, line 2 *Aspergillus fellis,* line 3 *Aspergillus luteonubrus,* line 4 *Aspergillus arizonicus,* and line 5 is *Rhizopus arrhizus*

This study concluded that different microorganism isolates survive and grow well in crude oil and were affiliated with *Bacillus anthracis*, *Bacillus cereus, Achromobacter sp.,* and *pseudomonas aeroginosa* for bacterial isolates. Moreover, *Rhizopus arrhizus* and *Aspergillus lentulus* for fungi isolates, this study showed these microorganisms have a strong ability to degrade crude oil and can be used to clean up soil and the environment. However, the longitudinal investigation will determine which enzyme is responsible for the degradation of crude petroleum and hydrocarbons, as well as whether the transcriptome encoding this enzyme is expressed by adding crude oil or if it is expressed simultaneously. And more studies are needed to identify other microorganisms in the Kurdistan Region with the ability to degrade crude oil, as well as to use a mixed culture of bacteria and fungi to observe their synergistic ability to degrade hydrocarbons.